\documentclass[reprint,prb,10pt,superscriptaddress,twocolumn,floatfix]{revtex4-1}

\usepackage{amsmath}  
\usepackage{amsfonts}
\usepackage{graphicx}   
\usepackage{epstopdf}	
\usepackage{verbatim}   
\usepackage{color}      
\usepackage{subfigure}  
\usepackage{hyperref}   
\usepackage{braket}
\usepackage{multirow}
\def\lmath#1{\text{\scalebox{1.3}{$#1$}}}
\def\lfrac#1#2{\lmath{\frac{#1}{#2}}}

\begin{document}
\label{mbs_readout}

\title{Readout of Majorana parity states using a quantum dot}

\author{Kaveh Gharavi}
\thanks{These authors contributed equally to this work.}
\affiliation{Institute for Quantum Computing, University of Waterloo, Waterloo, Ontario N2L 3G1, Canada}
\affiliation{Department of Physics and Astronomy, University of Waterloo, Waterloo, Ontario N2L 3G1, Canada}
\author{Darryl Hoving}
\thanks{These authors contributed equally to this work.}
\affiliation{Institute for Quantum Computing, University of Waterloo, Waterloo, Ontario N2L 3G1, Canada}
\affiliation{Department of Physics and Astronomy, University of Waterloo, Waterloo, Ontario N2L 3G1, Canada}
\author{Jonathan Baugh}
\email{baugh@uwaterloo.ca}
\affiliation{Institute for Quantum Computing, University of Waterloo, Waterloo, Ontario N2L 3G1, Canada}
\affiliation{Department of Physics and Astronomy, University of Waterloo, Waterloo, Ontario N2L 3G1, Canada}
\affiliation{Department of Chemistry, University of Waterloo, Waterloo, Ontario N2L 3G1, Canada}
\affiliation{Waterloo Institute for Nanotechnology, University of Waterloo, Waterloo, Ontario N2L 3G1, Canada}

\date{\today}
\pacs{73.63.Kv, 73.63.Nm, 74.45.+c, 74.78.Na, 85.35.-p}
\begin{abstract}
We theoretically examine a scheme for projectively reading out the parity state of a pair of Majorana bound states (MBS) using a tunnel coupled quantum dot. The dot is coupled to one end of the topological wire but isolated from any reservoir, and is capacitively coupled to a charge sensor for measurement. The combined parity of the MBS-dot system is conserved and charge transfer between the MBS and dot only occurs through resonant tunnelling. Resonance is controlled by the dot potential through a local gate and by the MBS energy splitting due to the overlap of the MBS pair wavefunctions. The latter splitting can be tuned from zero (topologically protected regime) to a finite value by gate-driven shortening of the topological wire. Simulations show that the oscillatory nature of the MBS splitting is not a fundamental obstacle to readout, but requires precise gate control of the MBS spatial position and dot potential. With experimentally realistic parameters, we find that high-fidelity parity readout is achievable given nanometer-scale spatial control of the MBS, and that there is a tradeoff between required precisions of temporal and spatial control. Use of the scheme to measure the MBS splitting versus separation would present a clear signature of topological order, and could be used to test the robustness of this order to spatial motion, a key requirement in certain schemes for scalable topological qubits. We show how the scheme can be extended to distinguish valid parity measurements from invalid ones due to gate calibration errors.  
\end{abstract}

\maketitle

\section{Introduction}
\label{sec:intro}
The elementary excitations of one-dimensional topological superconductors are Majorana Bound States (MBS), equal to their own anti-particles. This was first discovered by Kitaev \cite{kitaev}, and has spurred enormous interest \cite{beenakker_search_for_mf, alicea_new_directions, leijnse_felnsberg_review, tewari_review, DasSarma_Nayak_review_2015, MF_from_ferromagnets_2014, MF_from_ferromagnets_2015, MBS_evolution_of_dos_2015, flensberg_alicea_arxiv_review_2015} from the condensed matter community in the fundamental properties of this novel phase of matter, as well as its potential applications in topological quantum computation (TQC) \cite{nayak_review, Wu_new_braiding, Ivanov_non_abelian}. One recipe for MBS involves a semiconducting nanowire with a strong spin-orbit coupling, with induced superconductivity due to proximity with an s-wave superconductor. With the application of an external magnetic field of appropriate direction and magnitude, a pair of MBS appear at the ends of the nanowire as edge modes \cite{lutchynPRL2010_theory,sauPRL2010_theory,sauZBA10,Oreg2010}. As the MBS are zero energy modes, the ground-state is 2-fold degenerate. Several reports have been made on experimental evidence \cite{Mourik25052012, Das2012_1, RokhinsonObs2012, DengLundObs12, finkUrbanaObs13, MarcusObs2013} for the existence of this type of MBS, although a complete picture of the physics of systems hosting MBS, including conclusive evidence of the topological nature of the observed ground states, remains out of reach as of yet.

For the purposes of TQC, the degenerate MBS edge modes can be labelled $|0\rangle, |1\rangle$ in the computational basis, according to the \emph{parity} of the many-body ground state, with $|0\rangle \, (|1\rangle)$ referring to an even (odd) number of electrons. A so-called topological gap protects these states from the environment, providing an intrinsic, hardware-level protection against decoherence \cite{Alicea2011}. A \emph{logical} Majorana qubit is defined as the joint state of two MBS pairs within a particular parity manifold \cite{bravyi_review}. We shall focus on a single MBS pair here, as readout of a logical qubit can be constructed from pair readouts. A bit-flip operation $|0\rangle \leftrightarrow |1\rangle$ can be performed by utilizing the unusual MBS property of non-Abelian anyonic statistics. This involves braiding (physically exchanging the positions of) the two particles. The details of braiding operations were explored in ref. \cite{Alicea2011}, where it was also shown that these operations, as implemented in a network of quantum wires, benefit from topological error protection. However, in order to obtain a universal set of operations, one needs to supplement braiding with a set of quantum gates that are not topologically protected \cite{bravyi_review, bravyi_kitaev_clifford_gates, nayak_ising_anyons}. Several proposals exist for achieving universality, such as bringing the MBS close together to break topological protection and applying phase gates \cite{Alicea2011, bravyi_review}, or coupling MBS with conventional qubits \cite{fluxQubitReadout10, Flensberg2011, flensberg_spin}.

Additional challenges facing the realization of TQC are state initialization and readout of the MBS parity states. Following the methodology of the $\nu = 5/2$ fractional quantum Hall system\cite{read_green_FQHE, Freedman_FQHE}, a creation/annihilation approach was suggested by Alicea et al. \cite{Alicea2011}, wherein a pair of MBS are created from the vacuum of the underlying quantum field, braided to perform computation, and then fused (annihilated) to create either vacuum or a finite energy quasiparticle (i.e. a Dirac fermion), depending on the parity state of the MBS. The extra quasiparticle can be detected by some form of charge measurement. There are also recent proposals for readout based on monitoring the current-phase relation of a Josephson junction hosting an MBS pair \cite{lutchynPRL2010_theory, Heck2011}, coupling MBS to flux \cite{Hassler2010} or transmon \cite{Hassler_2011, Hyart_2013, flensberg_alicea_arxiv_review_2015} qubits, and coupling to charge or spin states of quantum dots \cite{Flensberg2011, topo_qbus,flensberg_spin}. All of these methods rely on some form of parity-to-charge conversion, and also necessarily take the MBS out of the topologically protected regime by breaking the degeneracy of the parity states. This can be achieved by reducing the spatial separation of the two MBS so their wavefunctions overlap \cite{Alicea2011}, or by using long-range Coulomb control interactions \cite{Hassler_2011, Hyart_2013} on a superconducting island hosting the MBS. Charge state coherence during the parity-to-charge conversion operation is generally required. 

In this paper, we propose and theoretically model a readout scheme that is relevant to the setup of MBS tunnel coupled to a quantum dot (QD). Previous theoretical work has demonstrated the power and versatility of the MBS-QD system for detecting the presence \cite{qdDetection11} and lifetime \cite{flensberg_lifetime} of topological order, gate-driven manipulation of topological qubits \cite{Flensberg2011, flensberg_spin, topo_qbus} and coherent transfer to dot spin and dot charge states. In a realization based on a top-gated two-dimensional electron gas (2DEG), for example, the MBS-QD setup is natural and could lead to a scalable architecture for topological qubits. While parity measurement was mentioned in the MBS-QD context \cite{Flensberg2011}, to our knowledge, no detailed study has been conducted to validate the experimental feasibility of such a readout scheme. Our setup involves an MBS pair, a QD isolated from any reservoir, and a charge sensor to measure the QD charge state. As there are no reservoirs present, the joint parity state of the QD + MBS system is conserved. By reducing the spatial separation of the two MBS (e.g. with a set of keyboard gates), the overlap of the MBS wavefunctions grow, resulting in an energy splitting between the $|0\rangle$ and $|1\rangle$ states. This splitting is oscillatory and has an exponential envelope versus the MBS separation \cite{kitaev, splitting_smoking_gun_12}. The QD level is tuned so that a charge transition is on resonance with a target MBS energy splitting, allowing MBS $\rightarrow$ QD charge transport to occur for one parity state but not the other. Similar to other schemes, we assume a coherent charge transfer process. Finally, the charge state of the QD is projectively measured with a charge sensor such as a single electron transistor (SET) \cite{Nilsson_2008}. 

Numerical simulations with realistic system parameters show that this setup can be used to map out the energy splitting between the $|0\rangle$ and $|1\rangle$ states versus the spatial separation of the MBS pair (or as a function of chemical potential or external magnetic field). Such a signature has been cited as ``smoking gun" evidence for topological order, and could also open avenues for studying the robustness of the topological state to domain wall motion. The charge transfer in our scheme can be performed on a fast timescale of $< 10$ nanoseconds with a high theoretical fidelity of $> 99\%$. These attributes can be further improved, but at a cost in the precision of voltage and timing controls. The isolation of the QD from reservoirs leads to a resonance in the tunneling probability versus gate voltage that is typically very sharp, and controlled only by the tunneling rate. While this requires some fine tuning of control parameters, it is very effective at decoupling the MBS and QD when readout is not being performed.

The manuscript is organized as follows: In section \ref{sec:model}, a model for the MBS pair coupled on one end to a QD is presented. In section \ref{sec:splitting}, we show how this setup can be used to experimentally determine the energy splitting between the MBS parity states as a function of their spatial separation. The MBS parity measurement is numerically studied and discussed in section \ref{sec:readout}, and concluding remarks are presented in section \ref{sec:conclusions}.

\section{Model}
\label{sec:model}
\begin{figure}[t]
	\centering
	\includegraphics[width=8.6cm]{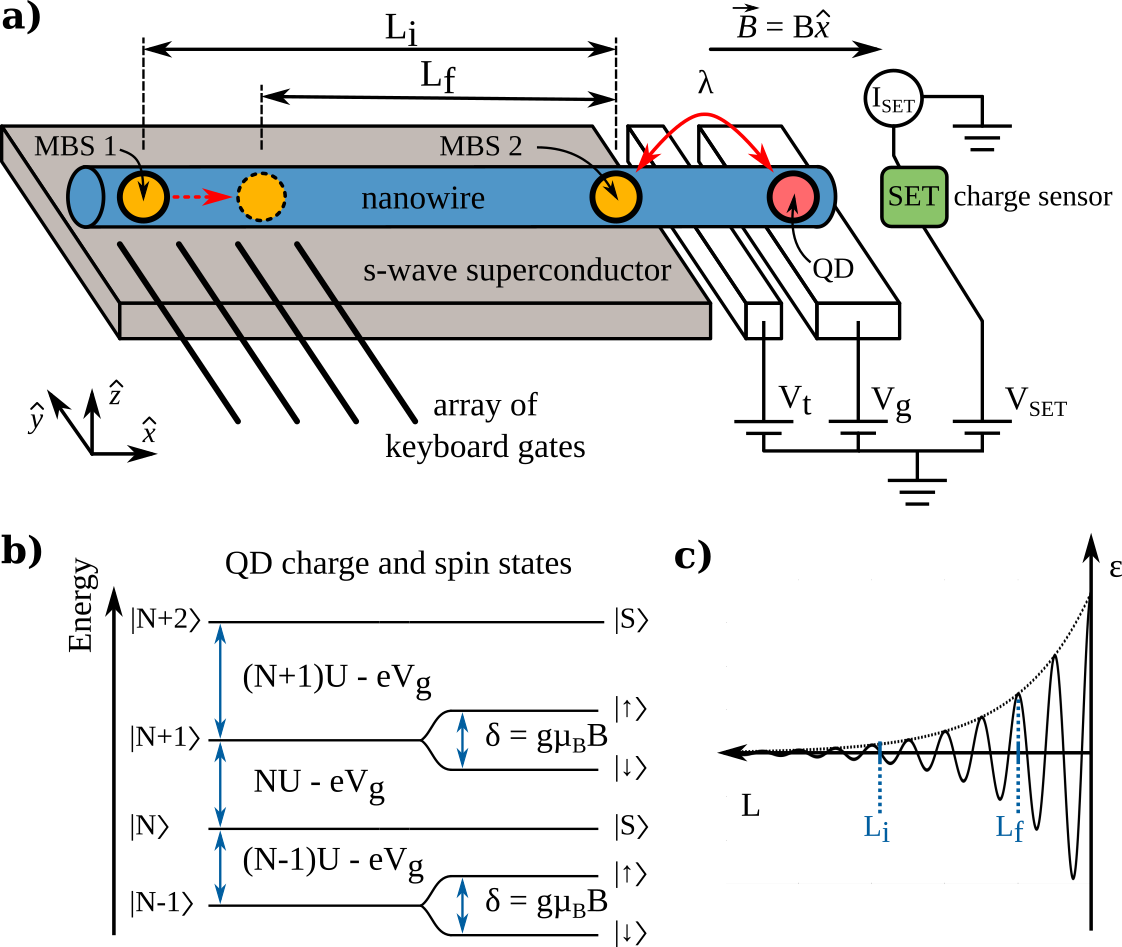}
	\caption{a) Schematic of the proposed device. A bulk s-wave superconductor is in close proximity to a semiconducting nanowire, inducing superconductivity in the nanowire. With the application of an axial (along $\hat{x}$) magnetic field, a pair of MBS appear at the ends of the topological region. An array of keyboard gates can be used to move MBS 1, tuning the MBS spatial separation from $L_i$ to $L_f$. MBS 2 is tunnel coupled to an isolated quantum dot (QD), with a tunneling strength $\lambda$ controlled by the gate voltage $V_t$. The chemical potential of the QD can be tuned using the plunger gate voltage $V_g$. A charge sensor reads out the charge state of the QD, shown here as a SET with current $I_{SET}$. b) Schematic of the energy levels of the QD. $U$ is the charging energy, and the charge state is indicated on the left by number of electrons on the QD. Integer $N$ is arbitrarily chosen to be even. Spin states are indicated on the right, with spin singlets (doublets) occurring for even (odd) charge states. A Zeeman splitting $\delta$ is induced between the spin-1/2 states by the external magnetic field. c) The qualitative behaviour of MBS energy splitting $\epsilon$ versus the separation $L$ is oscillatory with an exponential envelope. 
	  \label{fig:schematic}}
\end{figure}

Figure \ref{fig:schematic}a schematically illustrates the proposed setup for the initialization/readout scheme of the MBS parity state. A semiconducting nanowire with a strong Rashba-type spin-orbit coupling \cite{Wimmer_2015, Nadj-Perge_Rashba_InSb_2012} is contacted by a bulk s-wave superconductor, resulting in proximity induced superconductivity in the nanowire. The application of an axial magnetic field $\vec{B} = B \hat{x}$ of appropriate magnitude results in a phase transition to the topological regime \cite{lutchynPRL2010_theory}, with a pair of MBS emerging at the edges of the topological region. Using an array of keyboard gates located near one end of the nanowire, the chemical potential in the nanowire can be manipulated to move the edge of the topological region \cite{aliceaPRB2010_theory, Alicea2011}, thus tuning the separation between the two MBS from an initial value $L_i$ to a final value $L_f$. The MBS at the other end of the nanowire is tunnel coupled to an isolated quantum dot (QD) defined inside the nanowire. The energy level of the QD is controlled by the plunger gate voltage $V_g$, and the strength of the tunnel coupling by $V_t$. In particular, $V_g$ can be tuned such that the energy required to change the electron number on the dot matches the energy splitting of the MBS, i.e. the resonant tunneling condition. A nearby charge sensor, e.g. a single-electron transistor (SET) or quantum point contact, couples capacitively to the QD. A measurement of the sensor current results in a projective measurement of the QD charge state on a measurement timescale $t_m$, typically microseconds \cite{andrea_spin_qubit_2014, marcus_rfqpc}, but as short as $\sim 400$ ns \cite{Petta2015}. Readout of the QD charge state is the last stage of the MBS parity readout procedure, and $t_m$ is assumed to be much longer than the timescale for QD $\leftrightarrow$ MBS resonant tunneling, so the back-action from the charge sensor on the tunneling process is assumed to be negligible.

The two MBS are described by normalized second-quantized operators $\gamma$, which follow the Majorana fermion rules $\gamma_i = \gamma_i^\dagger$ and $\gamma_i ^2 = 1$ for $i = 1 , 2$. From these, we define a non-local Dirac fermion, with annihilation/creation operators $f = (\gamma_1 + i \gamma_2)/2$, $f^\dagger = (\gamma_1 - i \gamma_2)/2$. The MBS parity state is encoded as a single fermionic mode $|m\rangle$, where $m \in \{0,1\}$ is the occupation number of the non-local Dirac fermion.

A charge state with $N$ electrons on the QD, $|N\rangle$, is associated with electrostatic energy $E_N$. For even $N$, electrons are paired and form the spin singlet state $|S\rangle$; for odd $N$ the excess electron gives an overall spin-up  $\left| \uparrow \right\rangle$ or spin-down $\left| \downarrow \right\rangle $ state. A Zeeman splitting is induced for odd $N$ by the applied magnetic field $\vec{B}$. Figure \ref{fig:schematic}b shows the QD energy level diagram.

Without loss of generality, let the QD ground state consist of an even number of electrons $n$. The minimal model of the system consists of three fermionic modes: one each for spin-up and spin-down excitation on the QD and one for the MBS parity state. The charge on the QD is restricted to $n$, $n+1$, or $n+2$ electrons, which is made possible with a suitable choice for the gate voltage $V_g$. The $n \leftrightarrow n+1$ charge transition of the dot is later brought into resonance with the MBS, and used for parity readout. Charge transitions to the $n-1, n+2$ states, however, are not resonant because of energy separations on the order of the Coulomb charging energy, a few meV. This justifies excluding the $n-1$ state from the model. The $n+2$ state corresponds to both spin modes on the QD being occupied and is therefore included in the model, but its occupation probability remains negligibly small. This minimal model describes the system with an eight-dimensional Hilbert space, which is sufficient to capture the relevant dynamics while also being small enough for efficient numerical simulation.

The basis states are represented by $|N,\sigma,m\rangle$ where, $N \in \{n,n+1,n+2\}$, $\sigma \in \{S,\uparrow,\downarrow\}$, $m \in \{0,1\}$. However, it must be kept in mind that only the spin singlet is allowed for $N = n, n+2$, while for $n+1$ the singlet is disallowed.

The Hamiltonian is composed of four terms: $\mathcal{H} = \mathcal{H}_q + \mathcal{H}_s + \mathcal{H}_m + \mathcal{H}_t$, where the first three terms are diagonal and represent the dot charge, dot spin, and MBS energies, and $\mathcal{H}_t$ represents the tunnel coupling between the QD and MBS, which can depend on the spins of both systems. The dot charge term is
$$\mathcal{H}_q|N\rangle = E_N|N\rangle ,$$
where the constant interaction \cite{Beenakker_constant_interaction_QD_1991} energy $E_N = -eV_gN+\frac{U}{2}N(N-1)$ is used. $V_g$ is the voltage on the plunger gate and $U$ is the Coulomb charging energy. The remaining terms are: 
\begin{align}
\mathcal{H}_s &= \frac{\delta}{2} (\ket{\uparrow}\bra{\uparrow} - \ket{\downarrow}\bra{\downarrow} ), \nonumber \\
\mathcal{H}_m &= \frac{\epsilon}{2} (f^{\dagger}f - \frac{1}{2}), \nonumber \\
\mathcal{H}_{t} &= [\lambda_\uparrow (d_{\uparrow} - d^{\dagger}_{\uparrow}) + \lambda_\downarrow (d_{\downarrow} + d^{\dagger}_{\downarrow}) ] (f^{\dagger} + f), \nonumber 
\end{align}
where $\delta = g \mu_B B$ is the Zeeman energy of the dot spin, $\epsilon$ is the MBS energy splitting (which depends on the MBS separation $L$), $d_{\sigma} (d_{\sigma}^{\dagger})$ annihilates (creates) an electron with spin $\sigma$ on the dot, $\lambda_\sigma$ is the strength of the spin-dependent dot-MBS tunnel coupling, and $f, f^\dagger$ describe the non-local fermion defined previously. A matrix representation of the $d_\sigma, f$ operators is given in the Supplemental Material.

The spin polarization direction of the MBS depends on the relative strengths of the spin-orbit field of the nanowire and the Zeeman energy due to the external magnetic field \cite{Oreg2010,Simon2012}. If dominated by the Zeeman energy due the axial magnetic field, the MBS spin will be polarized along the $\pm \hat{x}$ (axial) direction. By contrast, for the spin-orbit dominated case, it will be polarized along the $\pm \hat{y}$ direction (in-plane, perpendicular to the nanowire axis). The MBS readout procedure is equally applicable to both cases, as explained below.

The QD-MBS tunnelling constant $\lambda_\sigma$ depends on the spins of both systems. An MBS spin along the $\pm\hat{x}$ direction is only coupled to one spin state on the dot. Specifically, $\lambda_\downarrow  = \lambda, \lambda_\uparrow = 0$ for the $-\hat{x}$ direction, and $\lambda_\uparrow  = \lambda,\lambda_\downarrow = 0$ for the $+\hat{x}$ direction. In contrast, an MBS spin along $\pm \hat{y}$ direction will couple to the two $\pm \hat{x}$ spins on the QD equally \cite{flensberg_spin}, e.g. $\lambda_{\uparrow} = \lambda / \sqrt{2}$ and $\lambda_{\downarrow} = -i\lambda / \sqrt{2}$ for the $-\hat{y}$ direction. For a generic MBS spin polarization (used below), $\lambda_\sigma$ will be in between these two limiting cases. Spin rotations induced by the nanowire spin-orbit interaction during the tunneling process are neglected: their effect is to give the tunneling spin a component along $\pm \hat z$, which can be captured by assuming an arbitrary MBS spin polarization.

The MBS splitting $\epsilon$ is proportional to the overlap of the MBS wavefunctions \cite{kitaev, splitting_smoking_gun_12}, which are localized at the edges of the topological region. The wavefunctions decay exponentially inside the topological region, with a characteristic length $\xi$ on the order of the phase coherence length inside the nanowire. For $L \gg \xi$, the parity states are sufficiently degenerate for topological protection of the system. As $L$ is shortened, the splitting oscillates within an exponentially increasing envelope, as described in ref. \cite{splitting_smoking_gun_12}. This is illustrated qualitatively in Figure \ref{fig:schematic}c. In the regime $L \gtrsim \xi$, ref. \cite{splitting_smoking_gun_12} gives the splitting as a function of $L$ as:
\begin{equation}\label{eq:epsilon_L}
\epsilon (L) \approx \hbar^2 \tilde{k}_F \frac{e^{-2L/\xi}}{m^* \xi} \cos{\tilde{k}_F L},
\end{equation}
where $\tilde{k}_F$ is the effective Fermi wave-vector of the MBS wavefuctions inside the nanowire, and $m^*$ is the effective electron mass. We show in the next section how a series of experiments can be used to map out $\epsilon(L)$. Precise knowledge of this function is required for the MBS parity readout scheme described in section \ref{sec:readout}.

\section{MBS energy splitting}
\label{sec:splitting}

In section \ref{sec:subsec_fixed_L}, we describe how to measure the MBS splitting $\epsilon$ at fixed $L$ using resonant tunneling with the QD. In section \ref{sec:subsec_varied_L}, $L$ is varied to show how the function $\epsilon(L)$ is mapped out. Parameters relevant to InSb nanowires are used throughout the paper, as listed in table \ref{table:1-static_variables}. The results of this paper do not depend strongly on the values of these parameters; rather they are chosen for their experimental relevance. We assume the quantum dot charging energy is $U = 5$ meV, and an effective superconducting gap of $\Delta = 0.5$ meV opens in the regions of the nanowire proximate to the superconductor. This value of $\Delta$ is chosen conservatively to pertain to experiments involving Nb, which has a superconducting gap of 1.4 meV. No sub-gap states (other than the two-fold degenerate MBS) are assumed to exist at energies below $\Delta$. An external axial magnetic field $\vec{B} = B \hat{x}$ of magnitude $B = 0.75$ T induces topological order in the superconducting section of the nanowire, where a chemical potential $\mu = 2$ meV is assumed. The spin-orbit energy in InSb nanowires is expected \cite{Wimmer_2015} to be in the range $0.25 - 1$ meV, smaller than the Zeeman splitting $\delta = 2.0$ meV at $B = 0.75$ T. A temperature $T = 50$ mK is used. Thus, the thermal energy $k_BT$ is is much smaller than the superconducting gap, $k_BT \ll \Delta$, and also the topological gap $k_BT \ll |\delta - \sqrt {\mu^2 + \Delta^2}|$. Under these conditions, the low energy states of the topological superconductor (i.e. the MBS) are well separated from all higher energy states, including the bulk superconducting states. The MBS are therefore isolated from the superconducting `lead'. As the quantum dot in our scheme is also isolated from metallic leads, we assume that temperature plays no role in the tunneling, which occurs between two isolated two-level systems. 

\begin{table}[t]
\centering

\begin{tabular}{cccc|cc|cccc}

$T$  & $B$  & $m^*$ 	& $g$  	& $k_B T$ 	& $\epsilon^*$ 			& $\Delta$ 	& $\delta$	& $\mu$		& $U$  	\\
(mK) & (T)  & ($m_e$)	& 		& \multicolumn{2}{c|} {($\mu$eV)} 	& \multicolumn{4}{c} {(meV)}				\\ \hline \hline
50   & 0.75 & 0.014		& 50 	& 4.3 		& $20 - 50$ 			& 0.5 		& 2.0 		& 2.0 		& 5.0                                 

\end{tabular}
\caption{Fixed parameters used throughout the paper, chosen based on their relevance to experiments on proximitized InSb nanowires. $T$ is the temperature (k$_B$ is Boltzmann's constant), $B$ the external axial magnetic field, $m^*$ the effective electron mass (in units of free electron mass $m_e$), and $g$ the Land\'{e} factor on the QD. Columns 5-10 show energies in ascending order: the thermal energy $k_B T$ is the lowest, followed by maximum MBS splitting $\epsilon^*$, proximity superconducting gap $\Delta$, Zeeman splitting $\delta$, chemical potential inside the InSb nanowire $\mu$, and QD charging energy $U$. The MBS parity readout procedure does not depend critically on these values, and is feasible over a large range of energy scales as long as the conditions $k_BT \ll \Delta$ and $k_BT \ll |\delta - \sqrt {\mu^2 + \Delta^2}|$ hold. \label{table:1-static_variables}}
\end{table}

\subsection{Fixed MBS separation}
\label{sec:subsec_fixed_L}
We fix the MBS pair separation so that the energy splitting $\epsilon$ at a value $\epsilon^*$ smaller than the (proximity) superconducting gap $\Delta$, hence the MBS do not couple to the continuum of quasi-particle states. The $|1\rangle , |0\rangle$ MBS parity states are then at energies $+\epsilon^{*} / 2, -\epsilon^{*}/2$ respectively. The gate voltage $V_g$ is tuned so that the number of electrons on the QD is $n$, as measured by the charge sensor. 

Consider an initial MBS parity state $|1\rangle$, so the initial state of the system is $|\psi_i\rangle = |n, S, 1\rangle$. The process $|n, S, 1\rangle \leftrightarrow |n+1, \sigma, 0\rangle$ is resonant when $\epsilon^*$ equals the energy cost $\Delta E_{n,\sigma}$ of the $|n,S\rangle \rightarrow |n+1, \sigma \rangle$ transition of the dot, with $\sigma = \uparrow$ or $\downarrow$. From the constant interaction model, we have $\Delta E_{n,\sigma} = -eV_g + nU \pm \delta/2$, where the Zeeman energy $\delta = g \mu_B B$ enters with a plus (minus) sign for $\sigma = \uparrow (\downarrow)$. Determining $\epsilon^*$ is based on finding the resonant gate voltage $V^*$. The value for the resonant gate voltage depends on the initial MBS parity state: had we started with the other parity state $|0\rangle$, both processes $|n, S, 0\rangle \leftrightarrow |n+1, \sigma, 1\rangle$ and $|n, S, 0\rangle \leftrightarrow |n-1, \sigma, 1\rangle$ would have been \emph{off-resonance} at the $V^*$ mentioned above. The first of the two processes is resonant at $V_g =  V^* + 2 \epsilon^*/e$ and the latter at $V_g = V^* + \delta/e - U/e$. This allows the MBS-dot setup to distinguish between the two MBS parity states.

Note that, due to Zeeman splitting of the spin levels of the QD, there are generally two possible values for $V^*$, labelled $V^{*}_\sigma$ for $\sigma = \uparrow, \downarrow$. Without loss of generality, we focus on the lower resonance voltage $V^*_\downarrow$ from this point onwards. Hence, we use the shorthand notation $\lambda$ to refer to $\lambda_\downarrow$, the tunnel coupling strength to the spin down state of the QD. For a generic MBS spin polarization direction, a second resonance voltage $V^*_\uparrow$ is present at $V^*_\uparrow = V^*_\downarrow + \delta/e$, but not used. The procedure can be readily extended to the special case of spin polarization along the $\pm \hat {x}$ axis, where only one resonant voltage is present: $V^*_\downarrow$ for the $- \hat x$ and $V^*_\uparrow$ for the $+ \hat x$ directions, respectively. We now turn our attention to finding $\epsilon^*$.

\begin{figure}[t]
	\centering
	\includegraphics[width=8.6cm]{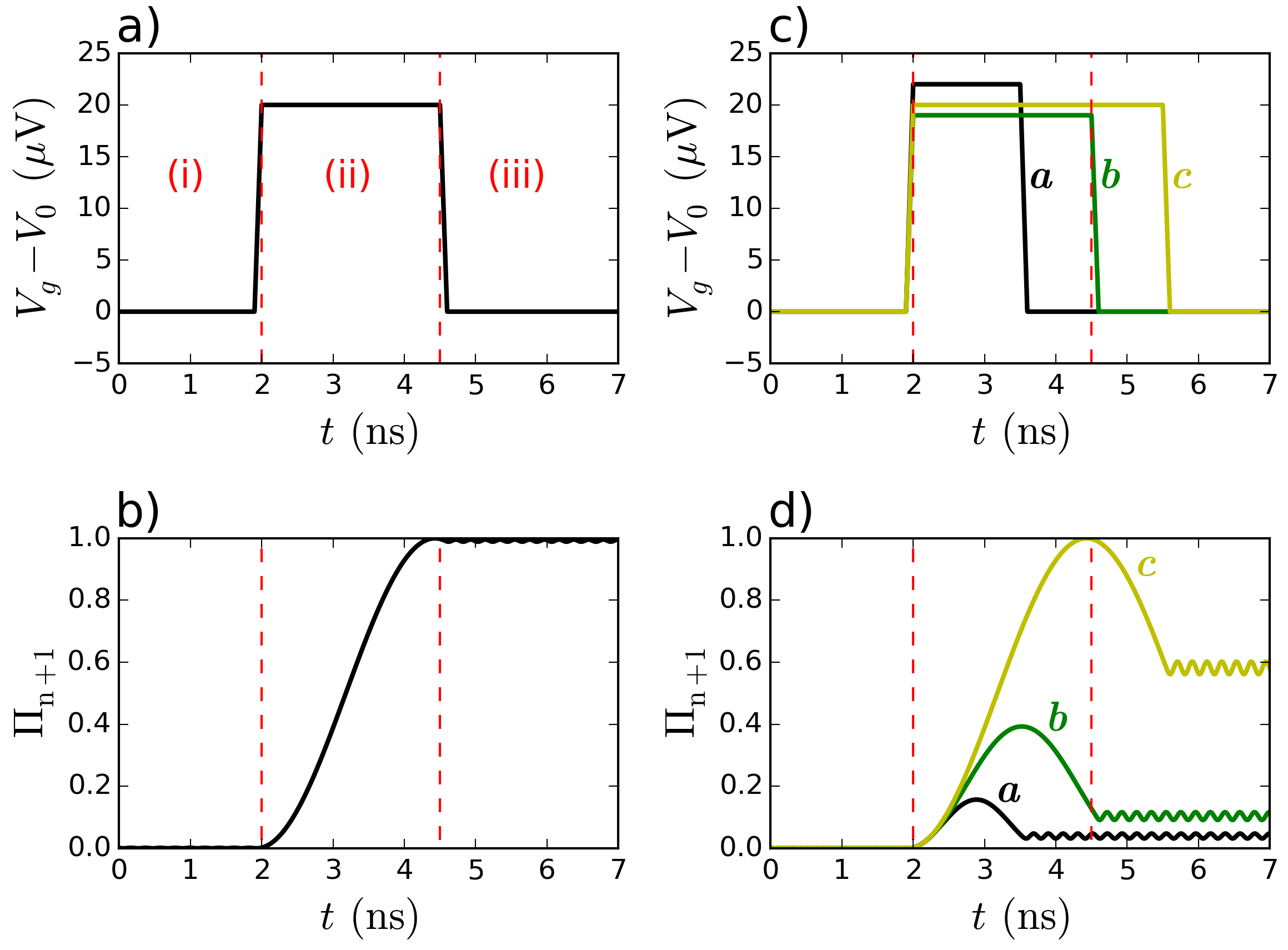}
	\caption{Procedure for determining $\epsilon$ for a fixed $L$: gate voltage $V_g$, and calculated probability of having $n+1$ electrons on the quantum dot, $\Pi_\mathrm{n+1}$, versus time. Panels a, b show the case in which the MBS and dot are brought into perfect resonance for the optimal charge transfer time.  a) $V_g$ is raised from the initial value $V_0$ to the resonance value $V^*_\downarrow = 20\, \mu$V, and held there for the optimal duration $T^* = 2.5$ ns, before being returned to $V_0$. The sequence is broken into three steps (i-iii). b) The corresponding probability $\Pi_{n+1}$ goes from zero to $> 99\% $. Panels c, d show three cases involving miscalibration of $V^*_\downarrow$ and $T^*$. c) Voltage sequences with $V^\mathrm{trial} = V^*_\downarrow +2\, \mu$V and $T = T^* - 1$ ns (curve `\textbf{\textit{a}}'), $T^{\mathrm{trial}} = T^*$ and $V^\mathrm{trial} = V^*_\downarrow -1\, \mu$V (curve `\textbf{\textit{b}}'), $V^\mathrm{trial} = V^*_\downarrow$ and $T^{\mathrm{trial}} = T^* +1$ ns (curve `\textbf{\textit{c}}'). d) Probabilities $\Pi_{n+1}$ corresponding to the sequences in panel c. In all panels, the vertical dashed lines show the optimal duration $T^*$ for resonant charge transfer. The following parameters are used: $\lambda/h = 100$ MHz,  $L = 1.12\, \mu$m, $\mu = 2$ meV, $B=0.75$ T. These correspond to $\epsilon^* = 20 ~\mu$eV.
	  \label{fig:optimize}}
\end{figure}

A procedure for determining $\epsilon^*$ is depicted in figure \ref{fig:optimize}, and is comprised of three steps: (i) The system starts in the state $|\psi_i \rangle$ with $V_g$ tuned to an initial value $V_0$, and $V_t$ at a large negative value so that tunneling between the MBS and QD is suppressed. At $t=0$, the tunnel coupling is turned on to a value $\lambda = h \times 100\; \mathrm{MHz}$ by tuning $V_t$. Then, at $t = 2$ ns, $V_g$ is rapidly ramped up to a trial value $V^\mathrm{trial}$, such that $\lambda^2 \ll \hbar e |d (V_g)/dt|$ at all times $t$, i.e. the state evolution is fast and non-adiabatic. This point is further discussed below. (ii) $V_g$ is held constant for the duration $T^\mathrm{trial}$, then (iii) rapidly ramped down to its initial value. The tunnel coupling is then turned off at $t = 7$ ns. Figure \ref{fig:optimize}a shows $V_g$ versus time, with ($V^\mathrm{trial}, T^\mathrm{trial}$) = ($V^*, T^*$), the values which produce resonant MBS-dot charge transfer for the chosen system parameters. The corresponding probability for charge transfer is shown in figure \ref{fig:optimize}b.\\

{\renewcommand{\arraystretch}{1.2}
\begin{table*}
\centering

\begin{tabular}{c|ccc|ccc}
 & $\lambda / h$ & $ d  (e V_g)/dt$ & $d (\epsilon) / dt$ 							 & 
	\multirow {2}{*}{$ \lfrac{(\Delta - \epsilon)^2} {\hbar d (e V_g)/dt}$ }     &
	\multirow {2}{*}{$ \lfrac{(\Delta - \epsilon)^2} {\hbar d (\epsilon) /dt}$ } &
	\multirow {2}{*}{$ \lfrac{\lambda^2} {\hbar d (e V_g)/dt}$ } \\
 & (MHz) 	& (meV/ns) 			& (meV/ns) \\ \hline \hline

Calibration & $100$  & $0.2$ &  --     &  $1.5 \times 10^{3}$  & --                  & $ 1.3 \times 10^{-3}$  \\                                     
Readout - 1 & $100$  & $1.3$ &  $0.8$  &  $2.3 \times 10^{2}$  & $3.8 \times 10^{2}$ & $ 2.0 \times 10^{-4}$  \\                                     
Readout - 2 & $1000$ & $1.3$ &  $0.8$  &  $2.3 \times 10^{2}$  & $3.8 \times 10^{2}$ & $ 2.0 \times 10^{-2}$  \\                                     

\end{tabular}
\caption{tunneling rate $\lambda / h$, maximum sweep rates of gate voltage energy $d (e V_g) / dt$ and MBS splitting energy $d (\epsilon) / dt$, and adiabaticity condition estimates, for three procedures discussed in the main text. ``Calibration" refers to procedure for finding $\epsilon^*(L)$ in section \ref{sec:subsec_fixed_L}; ``Readout - 1" refers to the MBS parity readout procedure discussed in detail in section \ref{sec:readout}, and ``Readout - 2" to the procedure at the end of section \ref{sec:readout} with $\lambda/h = 1$ GHz. Columns $4-6$ show unitless quantities comparing the gaps in the system's energy spectrum to the sweep rates of $V_g$ and $\epsilon$. A large number indicates a low probability of transition across the energy gap, whereas as a small number indicates a high transition probability. This probability can be roughly estimated from the Landau-Zener formula $\mathrm{exp} (-\alpha)$, where $\alpha$ refers to the table entries. Columns 4, 5: The probability of excitation of the MBS to a state within the continuum of states above $\Delta$ is expected to be negligibly small, i.e. the sweeps rates given for $V_g$ and $\epsilon$ are well within the adiabatic regime. This is confirmed in our numerical TDSE simulations. Column 6: The sweep rate of $V_g$ is fast compared to the tunnel repulsion $\lambda$ of the anti-crossing resonant states $|n,S,1\rangle, |n+1,\downarrow,0\rangle$, allowing an equal superposition to form with high probability. Thus, Rabi oscillations can occur as described in section \ref{sec:subsec_fixed_L}.
\label{table:2-rates}}
\end{table*}
}

\emph{Rapid sweep of $V_g$ and Rabi oscillations} -- Let us explore the resonant state transfer process (figure \ref{fig:optimize}a) in more detail. Starting at $V_0$, $V_g$ is swept to $V^{*}_\downarrow$. At this gate voltage, the states $|n,S,1\rangle$ and $|n+1, \downarrow, 0\rangle$ anti-cross due to the tunnel coupling $\lambda$. Note that $V_g$ is swept rapidly compared to the level repulsion $\lambda$, i.e. $\lambda^2 \ll \hbar e |d (V_g)/dt|$; however, it is swept \textit{adiabatically} slowly with respect to the continuum of states above the proximity gap: $|\Delta - \epsilon|^2 \gg \hbar e |d (V_g) / d t|$. Therefore, the probability of exciting to higher energy states is negligibly small. This is shown quantitatively in table \ref{table:2-rates}.

At the anti-crossing point, the eigenstates of the system are $|\pm\rangle = (\sqrt{2})^{-1} (|n,S,1\rangle \pm |n+1,\downarrow,0\rangle)$. However, since $V_g$ was swept rapidly, the system stays in its initial state $|\psi_i \rangle = |n,S,1 \rangle = (\sqrt{2})^{-1}(|+\rangle + |-\rangle)$. A Rabi oscillation occurs in the $\left\lbrace |+\rangle, |-\rangle \right\rbrace$ subspace, and after time $T^*$ the state of the system is $(\sqrt{2})^{-1}(|+\rangle - |-\rangle) =$ $|n+1,\downarrow,0\rangle$, up to an unimportant global phase. The system stays in this state after a rapid sweep of $V_g$ away from the anti-crossing point. Figure \ref{fig:optimize}b shows the simulated outcome of this process, obtained by numerically solving the time-dependent Schr{\"o}dinger equation (TDSE) to find $|\psi (t)\rangle$, the system state at time $t$. The quantity of interest is the probability of finding the dot in the $n+1$ charge state (with either spin), $\Pi_\mathrm{n+1}  (t) = \sum _{\sigma = \uparrow, \downarrow} |\langle n+1, \sigma, 0 |\psi (t)\rangle|^2$. It can be seen that $\Pi_\mathrm{n+1}$ goes from zero to $> 99 \%$.

By comparison, panels c,d of figure \ref{fig:optimize} pertain to the case of off-resonance charge transfer. For the same value of the initial gate voltage $V_0$ as in panel a, panel c shows $V_g$ versus time when $V^\mathrm{trial} = V^*_\downarrow +2\; \mu$V and $T = T^* - 1$ ns (curve `\textbf{\textit{a}}'), $T^{\mathrm{trial}} = T^*$ and $V^\mathrm{trial} = V^*_\downarrow -1\; \mu$V (curve `\textbf{\textit{b}}'), $V^\mathrm{trial} = V^*_\downarrow$ and $T^{\mathrm{trial}} = T^* +1$ ns (curve `\textbf{\textit{c}}'). The corresponding $\Pi_{\mathrm{n+1}}$ values are shown in figure \ref{fig:optimize}d, and indicate significant decreases compared to figure \ref{fig:optimize}b. The results indicate that the precision required for external control of voltage and time should be at the $100$ nV and $100$ ps levels, respectively, for a transfer probability close to 1. Both requirements can be satisfied with current technologies.

A small ripple oscillation can be seen in figures \ref{fig:optimize}b and \ref{fig:optimize}d. This is due to a finite off-resonant dot-MBS coupling when the initial voltage $V_0$ is not very far from the resonant voltage $V^*_\downarrow$. In figures \ref{fig:optimize}b and \ref{fig:optimize}d, the $V^*_\downarrow-V_0$ is only $20\; \mu$V. In section \ref{sec:readout}, we use a much larger value $\sim 1.3\; \mathrm{mV}$ for this difference and find that the ripple is no longer observed.\\

\emph{Measurement of $\epsilon^*$} -- To determine $V^*_\downarrow$ and $T^*$, one would repeat the sequence (i-iii) many times for each set of trial input parameters, each time measuring the charge state of the dot using the SET after step (iii). The frequency of the $|n+1\rangle$ outcomes yields an estimate of $\Pi_{n+1}$. The parameter space $(V,T)$ is then surveyed to find the resonant tunneling time $T^* = h/\lambda$ and resonant gate voltage $V^{*}_\downarrow$. The MBS splitting is given by
\begin{equation*}
\epsilon^* = -eV^{*}_\downarrow + nU - \delta/2.
\end{equation*}

\emph{Mixture of parity states} -- The calibration procedure as described here assumes the ability to reliably prepare the MBS in a particular parity state. Suppose, instead, that one can only prepare the MBS in a statistical mixture $\rho = p|0\rangle\langle0|+(1-p)|1\rangle\langle1|$. Then, due to the sharp dependence of transition probability on $V_g$, the procedure is still effective at measuring $\epsilon^*$. We note that if $V^*_\downarrow$ is the resonant gate voltage for the $|1\rangle \rightarrow |0\rangle$ parity transition process, then the $|0\rangle \rightarrow |1\rangle$ process will be resonant at $V_g = V^*_\downarrow + 2\epsilon^*/e$. Thus, one would observe two peaks in $\Pi_{n+1} (t)$ of height $p$ and $1-p$, separated by $2\epsilon^*$ along the $V_g$-axis. Peaks corresponding to the spin-up state of the QD will generally be visible as well (for MBS spin polarization not along $\hat{x}$), at $V^*_\uparrow = V^*_\downarrow + \delta/e$.

\subsection{Energy splitting versus MBS separation}
\label{sec:subsec_varied_L}
The procedure outlined in the previous section may be repeated for a variety of $L$ values using the keyboard gates, thereby allowing the experimenter to map out the oscillatory function $\epsilon(L)$. In the Supplemental Material, we estimate a typical spatial period of the oscillations of $\epsilon$ to be $\sim 30 \; \mathrm{nm}$. Therefore, reliably varying $\epsilon$ with a precision $\sim 100\; \mathrm{neV}$ requires tuning $L$ (e.g. using keyboard gates) with a precision at the $\sim$ 1 nm level. 

Empirical measurement of the function $\epsilon(L)$ is itself desirable, as it is a direct test of the validity of Eq. \eqref{eq:epsilon_L} and would be strong evidence for the non-local nature of the MBS wavefunctions and the presence of topological order. The search over the $(V,T)$ parameter space at each $L$ point can be speeded up by noting that $T^*$ depends only on the tunnel coupling strength $\lambda$ (Supplementary Material), which can be assumed constant, reducing the optimization to a one dimensional search for $V^*_\downarrow$ once $T^*$ is known. 
 
Along with the dependence of $\epsilon$ on MBS separation, the dependence of $\epsilon$ on other physical parameters such as the strength of the Zeeman field and the chemical potential may be mapped out. Although only the $L$-dependence is required for our proposed read out scheme, the model for the MBS system described in ref. \cite{splitting_smoking_gun_12} may be empirically tested with respect to several independent variables. Below, we describe how knowledge of the function $\epsilon(L)$ may be used for readout of the MBS parity state.

\section{Parity readout}
\label{sec:readout}
\emph{Initial state} -- The keyboard gates separate the two MBS by $L_i = 5\, \mu$m where the two parity states are degenerate to within $0.5\; \mu \mathrm{eV} \ll k_BT \simeq 4.3\; \mu$eV, given the parameters we have chosen. From data collected by the calibration procedure in section \ref{sec:subsec_varied_L}, a target readout length $L_f$ for the topological wire is chosen. At $L_f$, the MBS splitting $\epsilon(L_f)$ is such that $\epsilon(L_f) > \epsilon(L)$ for all $L > L_f$, so $L_f$ corresponds to a local peak of the function $\epsilon(L)$. For the numerical calculation of the TDSE, we choose $L_f = 0.775\, \mu$m, resulting in $\epsilon(L_f) = 49\, \mu$eV. The optimal gate voltage $V^{*}_\downarrow$ at $L_f$ for resonance with the spin-down dot state is assumed to be known, based on the calibration procedure above. Since $L_f$ corresponds to a peak in $\epsilon (L)$, resonance with the dot does not occur for $L > L_f$. The dot is initially in the $|n,S\rangle$ state, where we have arbitrarily chosen $n = 20$. The gate voltage $V_g$ is initially held at a value $V_0 = (1/e)U(n-1/2)$, halfway between the $(n+1)\leftrightarrow n$ and $n\leftrightarrow(n-1)$ charge degeneracy points of the QD, so $V^{*}_\downarrow - V_0 =(1/e)( U/2 - \delta/2 - \epsilon) \simeq 1.3$ mV. To restrict the dot to the $\{ |n\rangle$, $|n+1\rangle \}$ charge states, it is necessary that $V_g$ is kept within the range $(n-1)U + \epsilon + \delta/2 < eV_g \leq eV^{*}_\downarrow = nU -\epsilon - \delta/2$ at all times. \\

\begin{figure}[t]
	\centering
	\includegraphics[width=8.6cm]{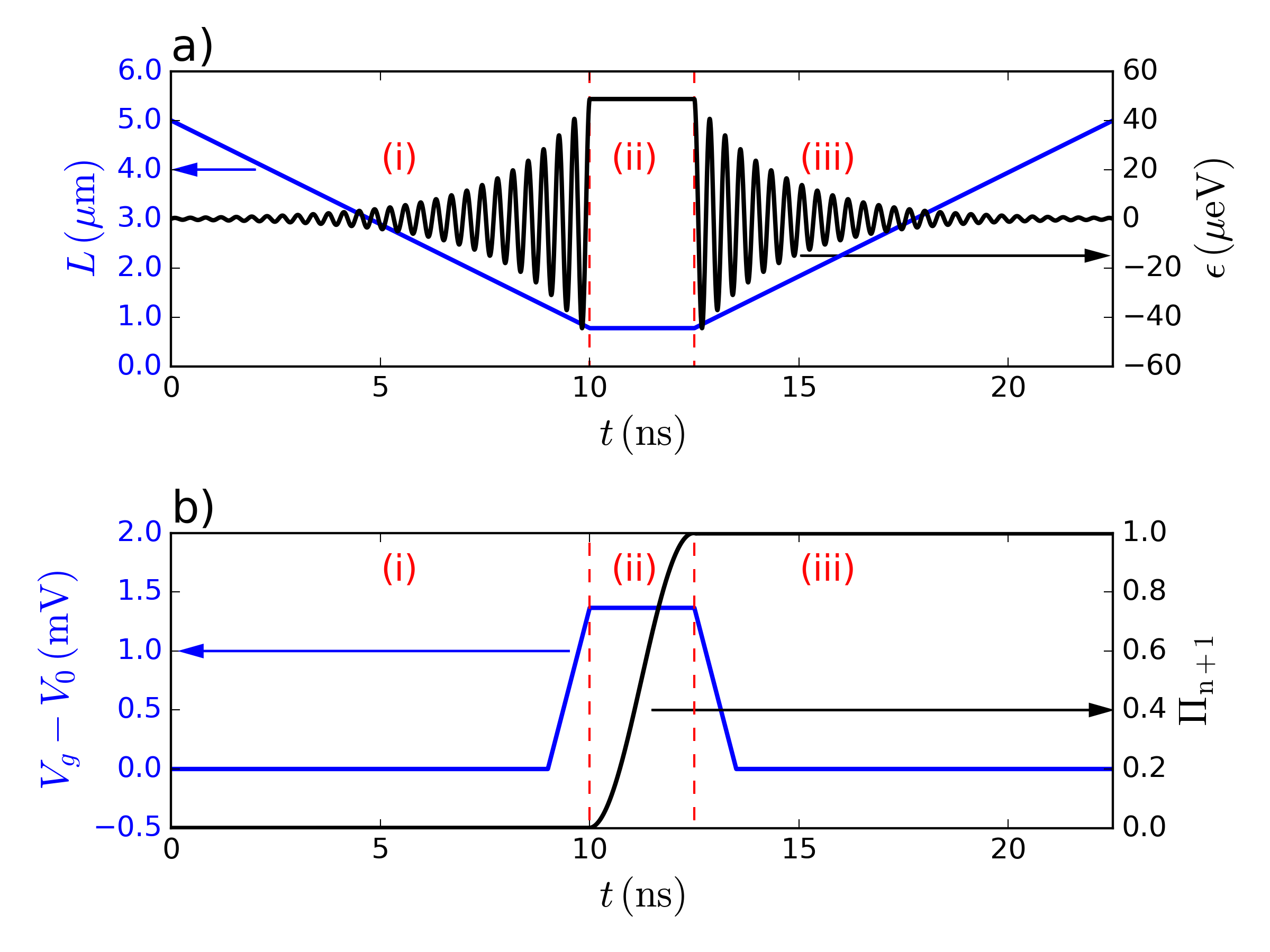}
	\caption{Readout procedure (stages i-iii) of the MBS parity state as described in the text. The MBS is initially in the $|1\rangle$ parity state. a) The MBS separation $L$ (blue/dark grey) and the corresponding MBS splitting $\epsilon (L)$ (black) as predicted from Eq. \ref{eq:epsilon_L}. b) The gate voltage $V_g$ (blue/dark grey) and the calculated probability of adding a charge to the dot, $\Pi_{n+1}$, versus time (black). The resonant gate voltage $V^*_\downarrow$ is known, obtained using the calibration procedure given in section \ref{sec:subsec_fixed_L}. As $V_g$ is tuned to $V^*_\downarrow$, the probability of finding $n+1$ electrons on the dot rises from zero to a value greater than 0.9999. Conversely, if MBS initial state is $|0\rangle$, the maximum $\Pi_{n+1}$ obtained is 0.004 (not shown). A charge readout of the dot then constitutes a readout of the MBS parity state. A tunneling strength  $\lambda/h = 100$ MHz is used in the numerical calculations.
	  \label{fig:readout}}
\end{figure}

\emph {MBS Parity readout procedure} -- With the system in its initial configuration, there are three stages of the read out, labelled (i), (ii), and (iii) in Fig. \ref{fig:readout}. In Fig. \ref{fig:readout}a, the MBS separation $L$, and the energy splitting $\epsilon (L)$ are shown as a function of time. Fig. \ref{fig:readout}b shows the gate voltage $V_g$, and the simulated probability $\Pi_\mathrm{n+1}  (t) = \sum _{\sigma = \uparrow, \downarrow} |\langle n+1, \sigma, 0 |\psi (t)\rangle|^2$, with the MBS initially in the $|1\rangle$ parity state. Considering each stage in turn:

Stage (i):
The keyboard gates move the left MBS towards the tunnel coupled end so that the MBS separation is reduced from $L_i = 5\, \mathrm{\mu m}$ to $L_f = 0.775\, \mathrm{\mu m}$. This is performed uniformly over a duration of 10 ns in our calculation. Table \ref{table:2-rates} shows that the adiabaticity condition  $|\Delta - \epsilon|^2 \gg \hbar |d \epsilon / d t|$ is satisfied at all times, so the probability of coupling to the continuum of quasi-particle states above and below $\Delta$ is negligible. Note that this step could be carried out much more slowly without affecting the results. 
The process of moving the left MBS can possibly incur dephasing errors within the $|0\rangle, |1\rangle$ parity basis. However, this does not adversely affect the readout procedure in any regard, as the readout is performed in the same parity basis. The parity eigenstates are \emph{preserved} under this transformation, as their levels cross but do not couple. At the end of this stage, $V_g$ controlling the dot potential is rapidly switched from $V_0$ to $V^{*}_\downarrow$. As discussed in table \ref{table:2-rates}, this transition is rapid with respect to $\lambda$ (so Rabi oscillation occur as explained in Section \ref{sec:subsec_fixed_L}), but adiabatic with respect to $|\Delta - \epsilon|$, so there is negligible chance of excitation to higher energy states. In our calculation the voltage ramping time is 1 ns.

Stage (ii):
The control parameters are held fixed for the optimal tunneling time $T^*$, which is $2.5$ ns in the case simulated here. With the MBS initially in the $|1\rangle$ state, the $|n,S,1\rangle \leftrightarrow$ $|n+1,\downarrow,0\rangle$ transition is on resonance, and an electron will tunnel from the topological wire to the dot with transition probability very close to one (Fig. \ref{fig:readout}b). 

If, however, the MBS was initially in the $|0\rangle$ state, changing the parity state will \emph{cost} (rather than supply) an energy $\epsilon(L_f)$. The corresponding process, $|n, S, 0\rangle \leftrightarrow |n+1, \downarrow, 1\rangle$, is off resonance -- its resonant gate voltage is $V_g =  V^{*}_\downarrow + 2 \epsilon/e$. For the $|0\rangle$ parity state then, the procedure illustrated in Fig. \ref{fig:readout} would result in an electron transfer probability very close to zero.

Stage (iii):
The reverse of stage (i), the gate voltage is rapidly ramped back to $V_0$ and the keyboard gates are used to move the left MBS back to its initial position. Note that, whereas sweeping $V_g$ away from the resonance point is necessary in order to prevent the electron from tunneling back to the MBS, moving the left MBS with the keyboard gates is not always required. It is included here to allow the system to recover its initial configuration, in case the cycle is repeated. At this point, a charge measurement of the dot is performed via the charge sensor, e.g. SET. A measurement outcome of $n+1$ indicates with high probability that the initial MBS state was $|1\rangle$ while a measurement of $n$ indicates with high probability that the initial MBS state was $|0\rangle$. Hence, the dot charge measurement amounts to a projective measurement in the MBS parity basis.\\

\emph{Fidelity of readout} -- Using the parameters given previously and with $\epsilon(L_f) = 49\; \mu$eV, the numerically obtained probability of finding $n+1$ electrons on the dot after stage (iii) is greater than $0.9996$ with the MBS initially in $|1\rangle$. The probability of finding $n$ electrons is greater than $0.9999$ with the MBS initially in $|0\rangle$. The readout scheme therefore allows the two MBS states to be distinguished with a visibility up to 0.9996, defined simply as the smaller of the two probabilities above. The term `readout fidelity' is used interchangeably with this measure of visibility. The residual error is dominated by the finite voltage ramping time: a faster ramp would increase the visibility. However, so far we haven't considered limitations on control precision (discussed below), which in practice lead to lower fidelities.  \\

\emph{Readout timescale} -- The timescale for the parity-dependent MBS $\rightarrow$ QD tunneling, including sweeping $V_g$ and moving the left MBS, can be as fast as 25 ns for experimentally feasible parameters (see figure \ref{fig:readout}). However, single-charge readout of the QD state requires integration times in the range of $0.4\; \mu$s \cite{Petta2015} to $10\; \mu$s \cite{marcus_rfqpc} or longer, and bottlenecks the MBS parity readout process.\\ 

\emph {Bias in parity readout due to miscalibration} -- Throughout the readout operation (stages i-iii), it was assumed that the calibration of $\epsilon(L)$ performed in section \ref{sec:subsec_varied_L} is valid. Drift or noise in the applied voltage or pulse timing will cause miscalibration errors and bias the charge measurement outcome in favour of $n$ over $n+1$ (see figure \ref{fig:optimize}), i.e. a bias towards detecting $|0\rangle$ over $|1 \rangle$ for the MBS parity. However, a straightforward modification of our scheme allows for distinguishing a calibration error from a genuine $|0\rangle$ outcome. This is done by appending a second readout operation involving the $n-1$ charge state of the QD.

Starting with $V_g$ at $V_0 = (1/e)U(n-1/2)$, i.e. halfway between the $(n+1)\leftrightarrow n$ and $n\leftrightarrow (n-1)$ charge degeneracy points of the QD, two parity-to-charge conversions are attempted: First, $|n, S, 1 \rangle \rightarrow |n+1, \downarrow, 0\rangle$, by using the resonance at gate voltage $V^{*}_\downarrow = V_0 + (1/e) ( U/2 -\delta/2 -\epsilon )$ as described previously. Subsequently, the $|n, S, 0 \rangle \rightarrow |n-1, \downarrow, 1\rangle$ transition is made resonant at $V_g = V_0 + (1/e) ( -U/2 +\delta/2 +\epsilon ) $. Then, the charge sensor is used to perform a charge readout of the QD. The following outcomes can be  distinguished: $n+1$ electrons indicates with high probability that the initial MBS state was $|1\rangle$, while $n-1$ indicates $|0\rangle$. The outcome $n$ indicates that neither transition took place (i.e. a calibration error), thus providing an \emph{in situ} test for the validity of the readout procedure.\\

\emph {Sensitivity to precision of control} -- For the system parameters chosen in our simulations, the MBS separation $L$ must be controlled within approximately 1 nm in order to maintain an accuracy $> 99\%$ in distinguishing the parity outcomes. The tolerance can be improved by about a factor of three by choosing parameters at the edge of the topological phase region that correspond to about three times longer period for the MBS energy oscillations -- however such a case is far less typical. Alternatively, the effect of tunnel broadening may be exploited to reduce the sharpness of the resonance condition and increase robustness. For example, we solved the TDSE again with a tunnel coupling strength of 1 GHz, corresponding to ``Readout - 2" in Table \ref{table:2-rates}. This shows that the stronger tunnel coupling allows a tolerance of $\pm 4$ nm in precision of the MBS location while still maintaining a readout fidelity of $\sim 97\%$, at the cost of reducing $T^*$ by a factor of 10. However, a $4$ nm error in the case of the 100 MHz tunnel coupling yields a dramatically lower visibility of $\sim 3 \%$. Hence, there is a tradeoff between the required precision of spatial control of the MBS separation versus the timing precision of gate voltage control. \\

\section{Conclusions}
\label{sec:conclusions}

We examined theoretically a protocol to read out the parity of an MBS pair in a topological superconductor using an isolated quantum dot. The MBS pair is brought from a well-separated (topologically protected) state to a spatially overlapping (unprotected) state in which there is a finite energy splitting; one MBS is then resonantly tunnel coupled with the quantum dot. The MBS parity state is projectively measured by a charge measurement of the quantum dot, and we showed that this can be accomplished, in principle, with high fidelity. It is straightforward to extend this to the readout of a logical qubit based on two MBS pairs. This protocol fits naturally into the MBS-dot system, which could be a powerful and versatile setting for achieving scalable control of topological qubits. 

As an intermediate step, we discussed a calibration procedure for mapping out the MBS energy splitting versus separation, $\epsilon(L)$. The result of such an experiment is predicted in ref. \cite{splitting_smoking_gun_12} and confirmation of this would be strong evidence for the presence of topological order. It would also allow testing the robustness of the MBS state against gate-driven motion of the topological domain wall. As with any projective measurement, the protocol can also be used to prepare the MBS into a desired parity eigenstate. The key for both readout and state preparation is that parity eigenstates should be preserved under adiabatic motion of the topological wire.\\

\begin{acknowledgements}

We thank K. Flensberg for helpful discussions. D. H. thanks M. Mosca for guidance and discussions. This work was supported by the Natural Sciences and Engineering Research Council of Canada and the Ontario Ministry for Research and Innovation.

\end{acknowledgements}


\begin{thebibliography}{10}

\bibitem{kitaev}
A.~Y. Kitaev,
\newblock Physics-Uspekhi {\bf 44}, 131 (2001).

\bibitem{beenakker_search_for_mf}
C.~Beenakker,
\newblock Annual Review of Condensed Matter Physics {\bf 4}, 113 (2013).

\bibitem{alicea_new_directions}
J.~Alicea,
\newblock Reports on Progress in Physics {\bf 75}, 076501 (2012).

\bibitem{leijnse_felnsberg_review}
M.~Leijnse and K.~Flensberg,
\newblock Semiconductor Science and Technology {\bf 27}, 124003 (2012).

\bibitem{tewari_review}
T.~D. Stanescu and S.~Tewari,
\newblock Journal of Physics: Condensed Matter {\bf 25}, 233201 (2013).

\bibitem{DasSarma_Nayak_review_2015}
S.~{Das Sarma}, M.~Freedman, and C.~Nayak,
\newblock Npj Quantum Information {\bf 1}, 15001 EP (2015).

\bibitem{MF_from_ferromagnets_2014}
J.~Li, H.~Chen, I.~K. Drozdov, A.~Yazdani, B.~A. Bernevig, and A.~H. MacDonald,
\newblock Phys. Rev. B {\bf 90}, 235433 (2014).

\bibitem{MF_from_ferromagnets_2015}
H.-Y. Hui, P.~M.~R. Brydon, J.~D. Sau, S.~Tewari, and S.~D. Sarma,
\newblock Scientific Reports {\bf 5}, 8880 EP (2015).

\bibitem{MBS_evolution_of_dos_2015}
T.~Kawakami and X.~Hu,
\newblock Phys. Rev. Lett. {\bf 115}, 177001 (2015).

\bibitem{flensberg_alicea_arxiv_review_2015}
D.~{Aasen}, M.~{Hell}, R.~V. {Mishmash}, A.~{Higginbotham}, J.~{Danon},
  M.~{Leijnse}, T.~S. {Jespersen}, J.~A. {Folk}, C.~M. {Marcus},
  K.~{Flensberg}, and J.~{Alicea},
\newblock ArXiv e-prints  (2015), 1511.05153.

\bibitem{nayak_review}
C.~Nayak, S.~H. Simon, A.~Stern, M.~Freedman, and S.~{Das Sarma},
\newblock Rev. Mod. Phys. {\bf 80}, 1083 (2008).

\bibitem{Wu_new_braiding}
L.-H. Wu, Q.-F. Liang, and X.~Hu,
\newblock Science and Technology of Advanced Materials {\bf 15}, 064402 (2014).

\bibitem{Ivanov_non_abelian}
D.~A. Ivanov,
\newblock Phys. Rev. Lett. {\bf 86}, 268 (2001).

\bibitem{lutchynPRL2010_theory}
R.~M. Lutchyn, J.~D. Sau, and S.~{Das Sarma},
\newblock Phys. Rev. Lett. {\bf 105}, 077001 (2010).

\bibitem{sauPRL2010_theory}
J.~D. Sau, R.~M. Lutchyn, S.~Tewari, and S.~{Das Sarma},
\newblock Phys. Rev. Lett. {\bf 104}, 040502 (2010).

\bibitem{sauZBA10}
J.~D. Sau, S.~Tewari, R.~M. Lutchyn, T.~D. Stanescu, and S.~{Das Sarma},
\newblock Phys. Rev. B {\bf 82}, 214509 (2010).

\bibitem{Oreg2010}
Y.~Oreg, G.~Refael, and F.~von Oppen,
\newblock Phys. Rev. Lett. {\bf 105}, 177002 (2010).

\bibitem{Mourik25052012}
V.~Mourik, K.~Zuo, S.~M. Frolov, S.~R. Plissard, E.~P. A.~M. Bakkers, and L.~P.
  Kouwenhoven,
\newblock Science {\bf 336}, 1003 (2012).

\bibitem{Das2012_1}
A.~Das, Y.~Ronen, Y.~Most, Y.~Oreg, M.~Heiblum, and H.~Shtrikman,
\newblock Nat Phys {\bf 8}, 887 (2012).

\bibitem{RokhinsonObs2012}
L.~P. Rokhinson, X.~Liu, and J.~K. Furdyna,
\newblock Nat Phys {\bf 8}, 795 (2012).

\bibitem{DengLundObs12}
M.~T. Deng, C.~L. Yu, G.~Y. Huang, M.~Larsson, P.~Caroff, and H.~Q. Xu,
\newblock Nano Letters {\bf 12}, 6414 (2012).

\bibitem{finkUrbanaObs13}
A.~D.~K. Finck, D.~J. {Van Harlingen}, P.~K. Mohseni, K.~Jung, and X.~Li,
\newblock Phys. Rev. Lett. {\bf 110}, 126406 (2013).

\bibitem{MarcusObs2013}
H.~O.~H. Churchill, V.~Fatemi, K.~Grove-Rasmussen, M.~T. Deng, P.~Caroff, H.~Q.
  Xu, and C.~M. Marcus,
\newblock Phys. Rev. B {\bf 87}, 241401 (2013).

\bibitem{Alicea2011}
J.~Alicea, Y.~Oreg, G.~Refael, F.~von Oppen, and M.~P.~A. Fisher,
\newblock Nat Phys {\bf 7}, 412 (2011).

\bibitem{bravyi_review}
S.~Bravyi,
\newblock Phys. Rev. A {\bf 73}, 042313 (2006).

\bibitem{bravyi_kitaev_clifford_gates}
S.~Bravyi and A.~Kitaev,
\newblock Phys. Rev. A {\bf 71}, 022316 (2005).

\bibitem{nayak_ising_anyons}
P.~Bonderson, D.~J. Clarke, C.~Nayak, and K.~Shtengel,
\newblock Phys. Rev. Lett. {\bf 104}, 180505 (2010).

\bibitem{fluxQubitReadout10}
F.~Hassler, A.~R. Akhmerov, C.-Y. Hou, and C.~W.~J. Beenakker,
\newblock New Journal of Physics {\bf 12}, 125002 (2010).

\bibitem{Flensberg2011}
K.~Flensberg,
\newblock Phys. Rev. Lett. {\bf 106}, 090503 (2011).

\bibitem{flensberg_spin}
M.~Leijnse and K.~Flensberg,
\newblock Phys. Rev. Lett. {\bf 107}, 210502 (2011).

\bibitem{read_green_FQHE}
N.~Read and D.~Green,
\newblock Phys. Rev. B {\bf 61}, 10267 (2000).

\bibitem{Freedman_FQHE}
M.~Freedman, C.~Nayak, and K.~Walker,
\newblock Phys. Rev. B {\bf 73}, 245307 (2006).

\bibitem{Heck2011}
B.~van Heck, F.~Hassler, A.~R. Akhmerov, and C.~W.~J. Beenakker,
\newblock Phys. Rev. B {\bf 84}, 180502 (2011).

\bibitem{Hassler2010}
F.~Hassler, A.~R. Akhmerov, C.-Y. Hou, and C.~W.~J. Beenakker,
\newblock New Journal of Physics {\bf 12}, 125002 (2010).

\bibitem{Hassler_2011}
F.~Hassler, A.~R. Akhmerov, and C.~W.~J. Beenakker,
\newblock New Journal of Physics {\bf 13}, 095004 (2011).

\bibitem{Hyart_2013}
T.~Hyart, B.~van Heck, I.~C. Fulga, M.~Burrello, A.~R. Akhmerov, and C.~W.~J.
  Beenakker,
\newblock Phys. Rev. B {\bf 88}, 035121 (2013).

\bibitem{topo_qbus}
P.~Bonderson and R.~M. Lutchyn,
\newblock Phys. Rev. Lett. {\bf 106}, 130505 (2011).

\bibitem{qdDetection11}
D.~E. Liu and H.~U. Baranger,
\newblock Phys. Rev. B {\bf 84}, 201308 (2011).

\bibitem{flensberg_lifetime}
M.~Leijnse and K.~Flensberg,
\newblock Phys. Rev. B {\bf 84}, 140501 (2011).

\bibitem{splitting_smoking_gun_12}
S.~{Das Sarma}, J.~D. Sau, and T.~D. Stanescu,
\newblock Phys. Rev. B {\bf 86}, 220506 (2012).

\bibitem{Nilsson_2008}
H.~A. Nilsson, T.~Duty, S.~Abay, C.~Wilson, J.~B. Wagner, C.~Thelander,
  P.~Delsing, and L.~Samuelson,
\newblock Nano Letters {\bf 8}, 872 (2008).

\bibitem{Wimmer_2015}
I.~van Weperen, B.~Tarasinski, D.~Eeltink, V.~S. Pribiag, S.~R. Plissard, E.~P.
  A.~M. Bakkers, L.~P. Kouwenhoven, and M.~Wimmer,
\newblock Phys. Rev. B {\bf 91}, 201413 (2015).

\bibitem{Nadj-Perge_Rashba_InSb_2012}
S.~Nadj-Perge, V.~S. Pribiag, J.~W.~G. van~den Berg, K.~Zuo, S.~R. Plissard,
  E.~P. A.~M. Bakkers, S.~M. Frolov, and L.~P. Kouwenhoven,
\newblock Phys. Rev. Lett. {\bf 108}, 166801 (2012).

\bibitem{aliceaPRB2010_theory}
J.~Alicea,
\newblock Phys. Rev. B {\bf 81}, 125318 (2010).

\bibitem{andrea_spin_qubit_2014}
M.~Veldhorst, J.~Hwang, C.~Yang, W.~Leenstra, B.~de~Ronde, J.~Dehollain,
  J.~Muhonen, F.~Hudson, K.~Itoh, A.~Morello, and A.~Dzurak,
\newblock Nat Nano {\bf 9}, 981 (2014).

\bibitem{marcus_rfqpc}
D.~J. Reilly, C.~M. Marcus, M.~P. Hanson, and A.~C. Gossard,
\newblock Applied Physics Letters {\bf 91} (2007).

\bibitem{Petta2015}
J.~Stehlik, Y.-Y. Liu, C.~M. Quintana, C.~Eichler, T.~R. Hartke, and J.~R.
  Petta,
\newblock Phys. Rev. Applied {\bf 4}, 014018 (2015).

\bibitem{Beenakker_constant_interaction_QD_1991}
C.~W.~J. Beenakker,
\newblock Phys. Rev. B {\bf 44}, 1646 (1991).

\bibitem{Simon2012}
D.~Sticlet, C.~Bena, and P.~Simon,
\newblock Phys. Rev. Lett. {\bf 108}, 096802 (2012).

\end{thebibliography}

\end{document}